\documentclass[a4paper,11pt]{article}
\usepackage{pos}
\usepackage{natbib}
\newcommand{\mus}{$\mu s$}

\title{Real-time Anomaly Detection at the L1 Trigger of CMS Experiment}

\author*[a]{Abhijith Gandrakota}

\onbehalf{for the CMS collaboration}

\affiliation[a]{Fermi National Accelerator Laboratory,\\
  Batavia, IL, USA}

\emailAdd{abhijith.gandrakota@cern.ch}

\abstract{We present the preparation, deployment, and testing of an autoencoder trained for unbiased detection of new physics signatures in the CMS experiment Global Trigger (GT) test crate FPGAs during LHC Run 3. The GT makes the final decision whether to readout or discard the data from each LHC collision, which occur at a rate of 40 MHz, within a 50 ns latency. The Neural Network makes a prediction for each event within these constraints, which can be used to select anomalous events for further analysis. The GT test crate is a copy of the main GT system, receiving the same input data, but whose output is not used to trigger the readout of CMS, providing a platform for thorough testing of new trigger algorithms on live data, but without interrupting data taking. We describe the methodology to achieve ultra low latency anomaly detection, and present the integration of the DNN into the GT test crate, as well as the monitoring, testing, and validation of the algorithm during proton collisions.}

\FullConference{%
  42$^{\text{nd}}$ International Conference on High Energy Physics - ICHEP 2024\\
  18-24 July 2024\\
  Prague, Czech Republic
}

\begin{document}
\maketitle

\section{Introduction}

The Large Hadron Collider (LHC), operating at the energy frontier, collides protons at an unprecedented rate of 40 million times per second (40 MHz). The Compact Muon Solenoid (CMS) experiment studies these collisions to uncover potential Beyond Standard Model (BSM) physics and precisely measure rare Standard Model (SM) processes \cite{cms}. While the high collision rate at the LHC increases the probability of producing and detecting rare processes, the nearly 100 million channels of the CMS detector also generate an enormous amount of data \cite{CMS:2020cmk,CMS:2016ngn}.

Only a small fraction of the 40 MHz proton-proton collision events—around 1,000 per second—can be stored for detailed offline analysis. To meet this stringent data reduction, events are selected using a two-tiered trigger system. The first level (L1), composed of custom hardware processors built with field-programmable gate arrays (FPGAs), uses information from the calorimeters and muon detectors to select events at a rate of around 100 kHz within a fixed latency of 4\mus~\cite{CMS:2020cmk}. The second level, the high-level trigger (HLT), consists of a processor farm running optimized event reconstruction software, reducing the rate to around 1 kHz before storage~\cite{CMS:2016ngn}. The L1 trigger discards the largest fraction of events, playing a crucial role in the two-tier trigger system. While this approach has enabled the discovery of the Higgs boson and many SM measurements, it risks missing unexpected new physics signatures not anticipated by predefined selection criteria.

This reliance on specific criteria introduces the risk of bias, potentially limiting the ability to detect new physics in the absence of strong theoretical predictions. This highlights the need for more general, unbiased triggering methods that maintain sensitivity to a wide range of new physics signatures. Ideally, such methods should identify anomalous events that deviate from expected SM patterns without relying heavily on specific BSM model predictions. For this reason, there are ongoing efforts to explore unsupervised machine learning (ML) techniques, such as anomaly detection (AD) methods, to select events with potential BSM signatures \cite{govorkova2022autoencoders}.

In this work, we present a novel approach: the implementation of ML-based real-time AD algorithms at the CMS L1 Trigger level. This cutting-edge technique provides a model-independent method for identifying potentially interesting events while maintaining high rejection rates for background processes. We discuss the development, implementation, and preliminary results of this approach, recently deployed at the CMS L1 Trigger \cite{dpn2024}. By introducing advanced ML capabilities at the earliest stage of event selection, CMS is entering a new era of data collection strategy, enhancing the search for unknown phenomena and potentially revolutionizing the approach to triggering in high-energy physics experiments.

\section{Anomaly detection L1 triggers in CMS}

The CMS collaboration has developed two complementary approaches to anomaly detection at the L1 trigger: AXOL1TL (Anomaly eXtraction Online Level-1 Trigger aLgorithm) and CICADA (Calorimeter Image Convolutional Anomaly Detection Algorithm). Both methods leverage advanced machine learning techniques to identify atypical events in real-time, offering a model-independent approach to capturing potential new physics signatures.

One of the well-known ML techniques for anomaly detection is the use of autoencoder (AE) \cite{ govorkova2022autoencoders}. Autoencoders are neural networks designed to learn efficient encodings of the dominant SM process in the input data. The architecture of an autoencoder consists of two main components: an encoder, which compresses the input data into a lower-dimensional representation (often called the latent space), and a decoder, which attempts to reconstruct the original input from this compressed representation.

These neural networks(NN) are trained to minimize the difference between the input and the reconstructed output. When trained on the proton-proton collisions as detected by the CMS detector, the network efficiently reconstructs the dominant background processes. When presented with an event containing processes BSM physics or rare SM physics; an anomalous event, the autoencoder struggles to reconstruct it accurately. This reconstruction error serves as our anomaly score; events with high reconstruction errors are flagged as potentially interesting.

Both the anomaly detection algorithms are trained on the trigger less stream, (ZeroBias dataset) collected by the CMS experiment in 2023 at a center-of-mass energy of $\sqrt{s}=13.6$ TeV, where lumi-leveling at pile-up 62 was used. 50\% of this used for training and 50\% was used for validation, performance testing. The anomaly detection triggers are estimated to have better performance than a traditional rule based trigger with similar event collection rate. In the case of AXOL1TL, it is to have 46\% efficiency gain when compared to rest of L1 trigger, when operating at a rate of $1 kHz$ for caputing exotic decay of higgs to four b quarks.

\subsection{Inputs to L1 anomaly detection }

\begin{figure} 
\centering 
\includegraphics[width=0.25\linewidth]{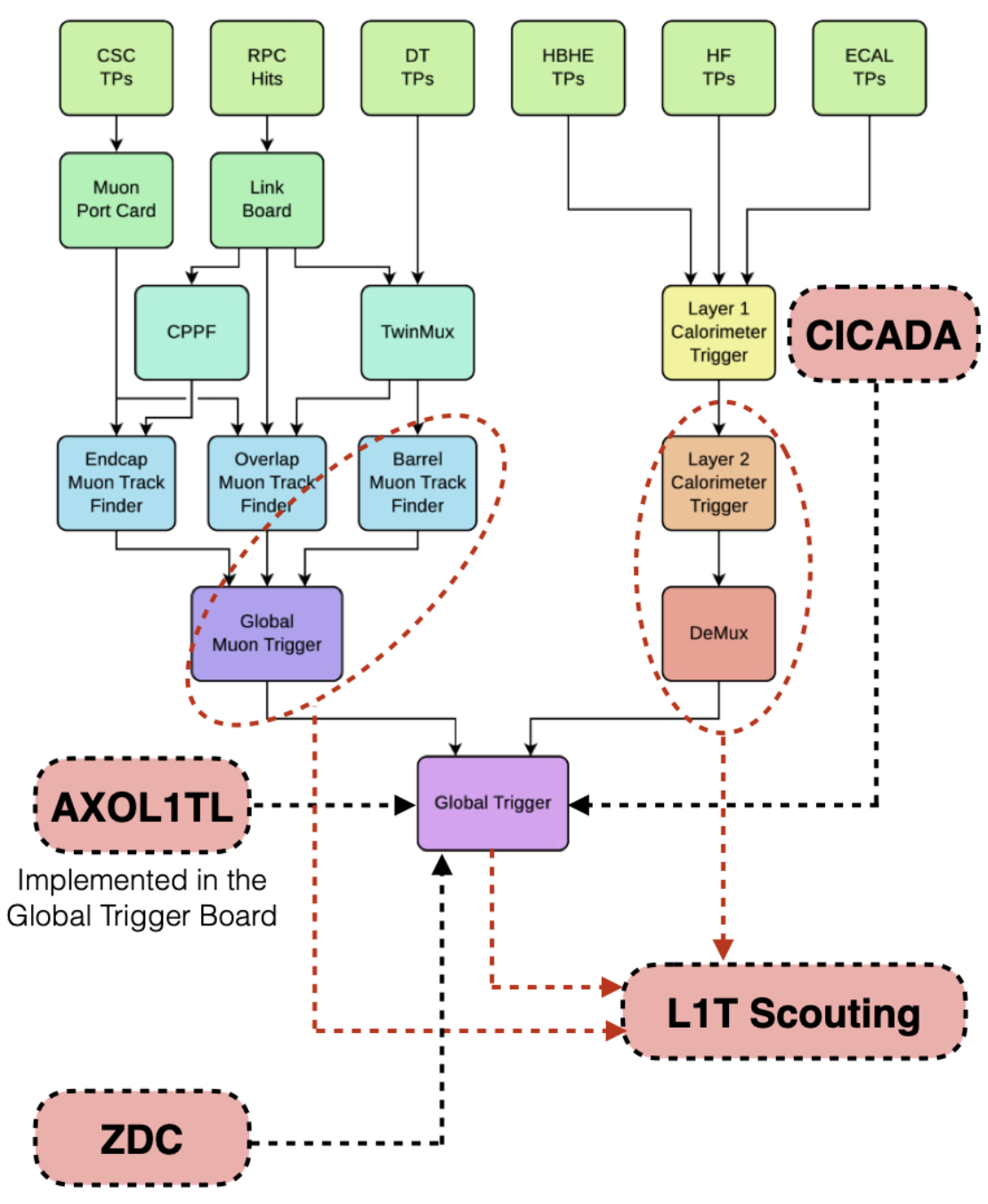} 
\caption{Components of the L1 trigger, along with the input and output paths for the anomaly detection algorithms.} \label{fig:l1t} 
\end{figure}

AXOL1TL and CICADA use different L1 reconstructions as inputs. AXOL1TL takes in 10 jets, 4 electron/photon objects, 4 muons, and transverse missing energy (MET) as reconstructed in the L1 trigger from calorimeter and muon triggers. The 3-momenta ($p_T$, $\eta$, $\phi$) of these objects, in raw hardware integer values, are used. CICADA, by contrast, uses calorimeter region energies, which form an 18x14 image-like input. The input streams for both algorithms are shown in Fig. \ref{fig:l1t}.

\subsection{Network Architecture and FPGA implementations}

AXOL1TL uses a variation of autoencoders called Variational Autoencoders (VAEs), which impose additional constraints on the latent space \cite{vae}. The encoder computes a latent space vector of Gaussian distributions $N(\mu=0,\sigma=1)$, with a latent vector size of 8. VAEs minimize both the reconstruction error and the deviation from a standard normal distribution by penalizing the KL divergence ($\mathcal{D}_{KL}$). The encoder and decoder are dense feed-forward networks. Only the encoder is used for real-time inference, and the anomaly score is approximated by the sum of squared means of the latent vector ($\sum_{i=1}^8\mu_i^2$).

CICADA uses convolutional layers in its autoencoder \cite{cnn}, as it processes image-like inputs of 252 (18x14) transverse energy deposits in two channels. The reconstruction quality is measured by Mean Squared Error (MSE) loss. To enable ultra-low latency inference, the autoencoder is shrunk using knowledge distillation to mimic the MSE loss with the given input \cite{hinton, ryan}.

\begin{figure} \centering 
\includegraphics[width=0.27\linewidth]{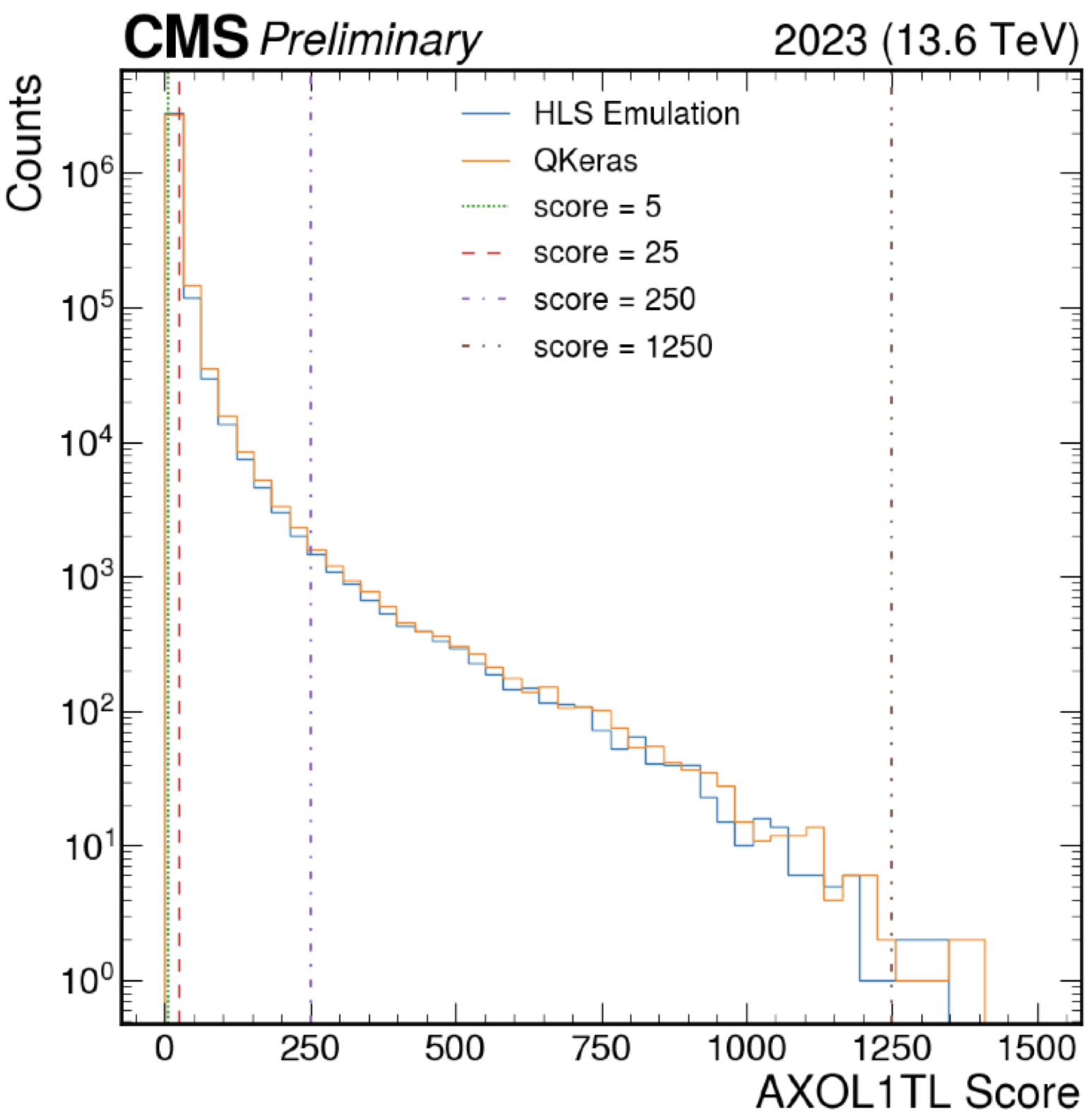} \includegraphics[width=0.37\linewidth]{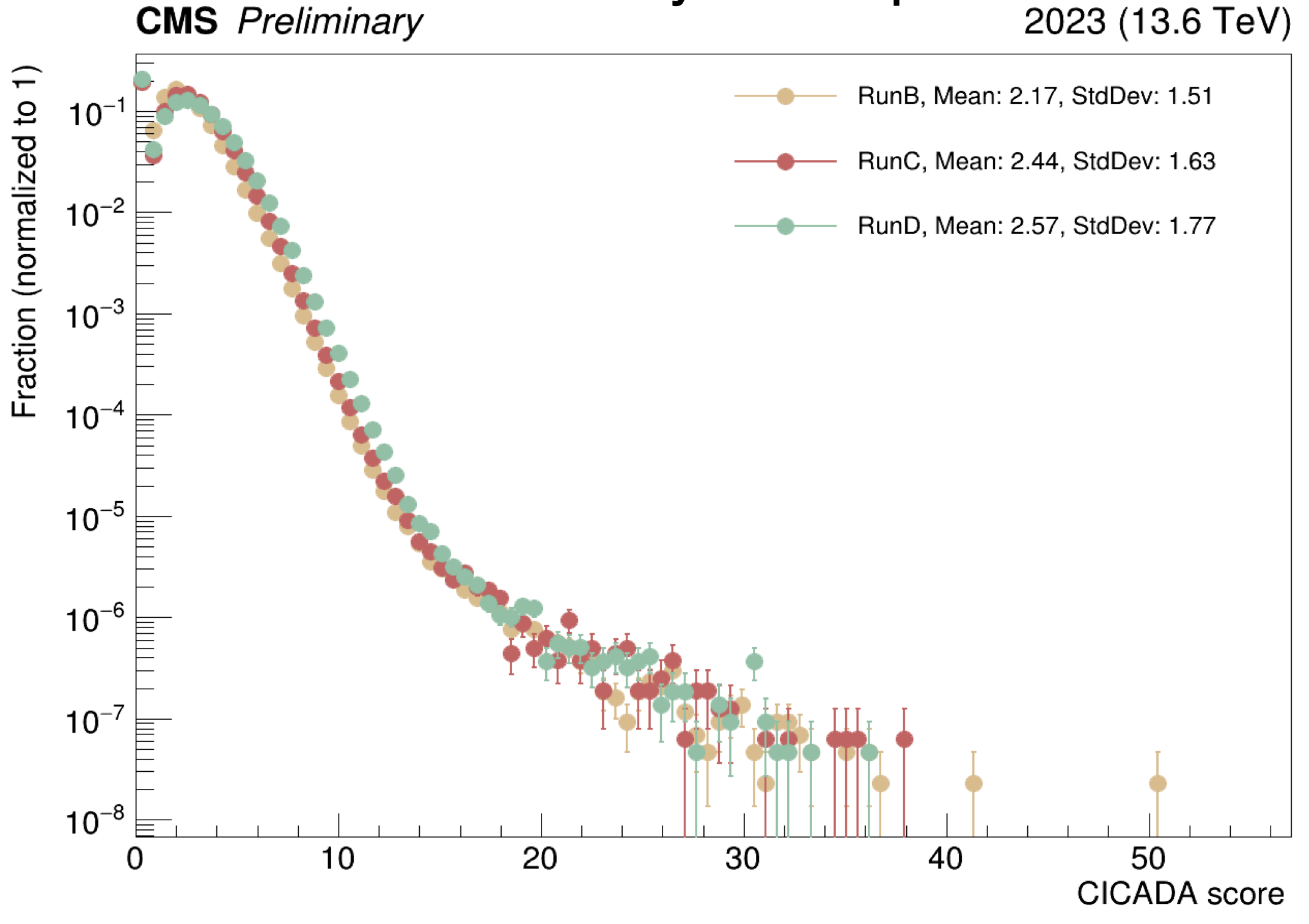} 
\caption{Anomaly score distributions from AXOL1TL (left) and CICADA (right) as output by both HLS emulation and qkeras. Higher-scoring events are flagged as anomalous for further analysis.} 
\label{fig:score} 
\end{figure}

Both algorithms are trained using quantization-aware training frameworks like qkeras \cite{Coelho_2021}. The networks are then translated to high-level synthesis (HLS) code using \textsc{hls4ml} \cite{Duarte:2018ite,fastml_hls4ml,Aarrestad:2021zos}, and deployed on Xilinx Virtex-7 FPGAs for ultra-fast inference. These FPGA implementations are carefully tuned for bit-exactness, as shown in Fig. \ref{fig:score}.

\section{Performance of AXOL1TL}
\begin{figure} \centering 
\includegraphics[width=0.62\linewidth]{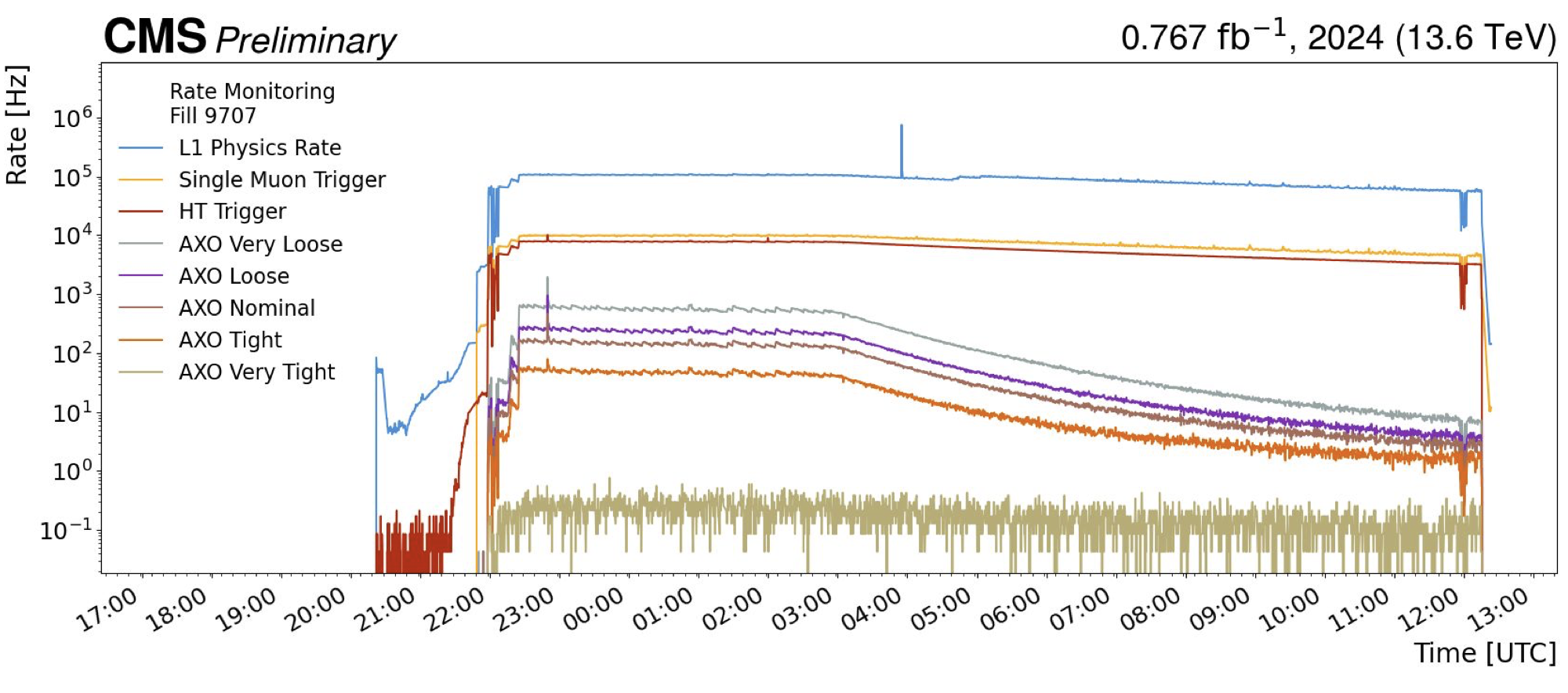} 
\caption{Global trigger rate monitoring time series over the course of data-taking in June 2024, showing the rates of AXOL1TL seeds.} 
\label{fig:rates} 
\end{figure}

Out of both anomaly detection triggers, AXOL1TL was commissioned for real-time testing and data-taking with the CMS trigger framework. The AXOL1TL trigger was tested at five different thresholds for triggering anomalous events in the data: very tight, tight, nominal, loose, and very loose. Events triggered through the nominal seed are sent for further reconstruction with HLT scouting and stored with minimal information \cite{scouting}. Events triggered by the very tight seeds are further processed for offline full event reconstruction. The AXOL1TL triggers performed stably during data-taking, as seen in Fig. \ref{fig:rates}.

\begin{figure} 
\centering 
\includegraphics[width=0.3\linewidth]{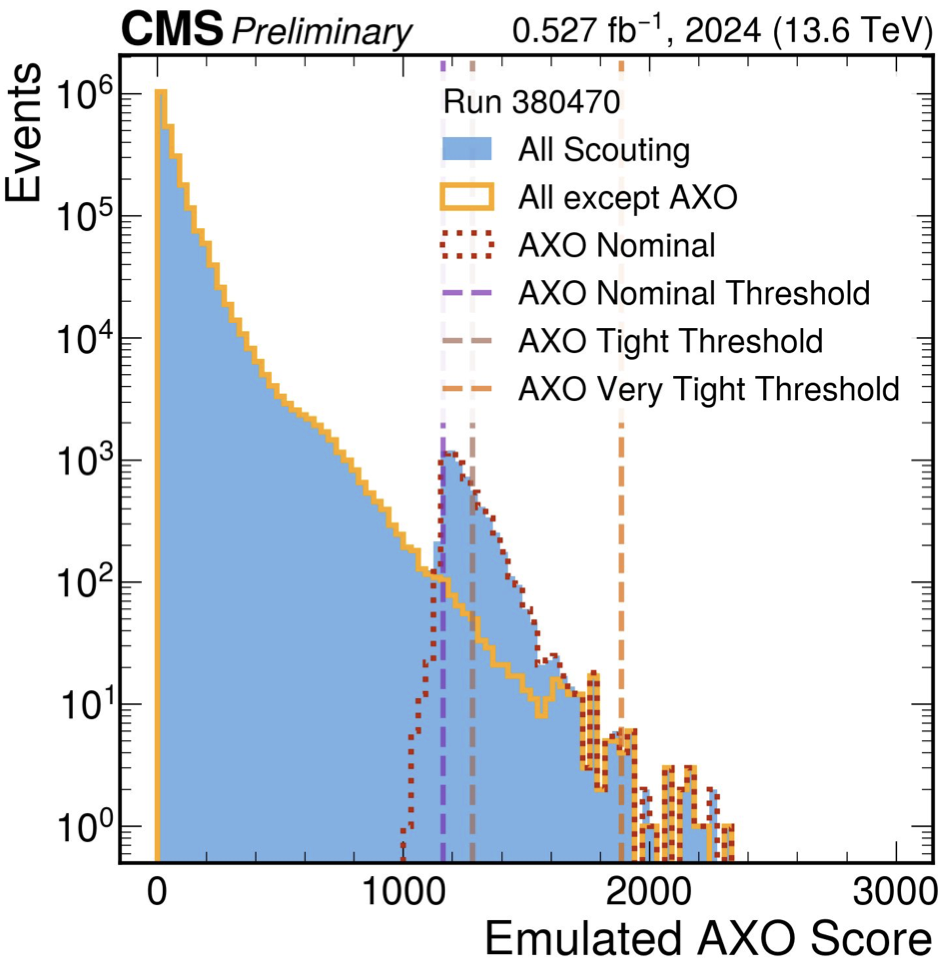} \includegraphics[width=0.3\linewidth]{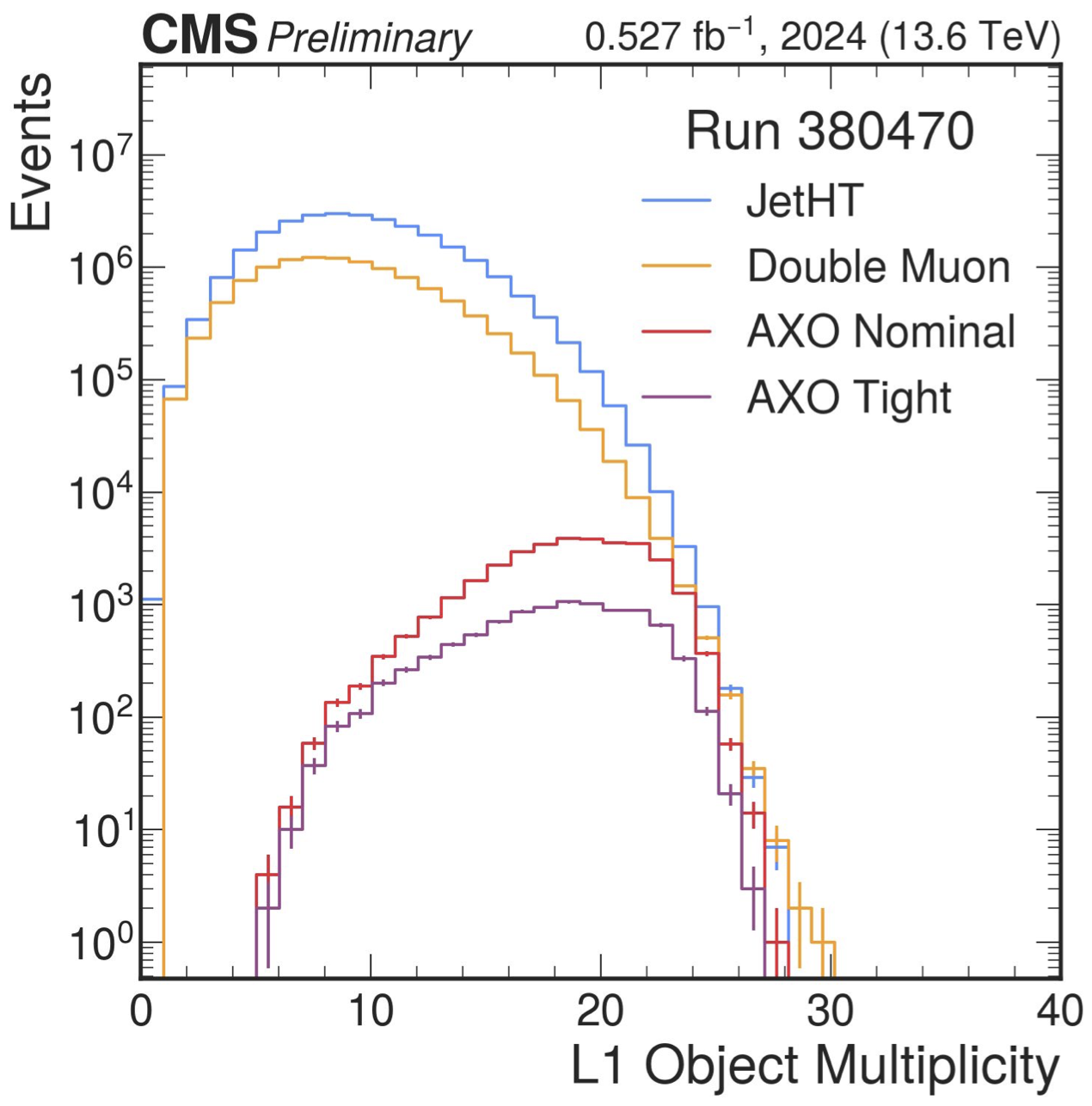} 
\caption{Left: Scores for all live AXOL1TL seeds and all events triggered by non-AXOL1TL HLT Scouting seeds, showing where the AXOL1TL contribution lies. Right: The distribution of AXOL1TL scores as a function of L1 object multiplicity.} 
\label{fig:events} 
\end{figure}

The dataset triggered by AXOL1TL tends to be orthogonal to events triggered by the regular L1 Trigger menu, as seen in Fig. \ref{fig:events}. This orthogonality highlights the novelty of the events collected by the L1 anomaly detection triggers. On closer examination of events triggered by both AXOL1TL and CICADA, we observe a preference for marking events with higher multiplicity as more anomalous. This reinforces the key role of these triggers in selecting events with multijet final states and BSM signatures, such as SUEPs, often overlooked by traditional rule-based algorithms.

\begin{figure} 
\centering 
\includegraphics[width=0.85\linewidth]{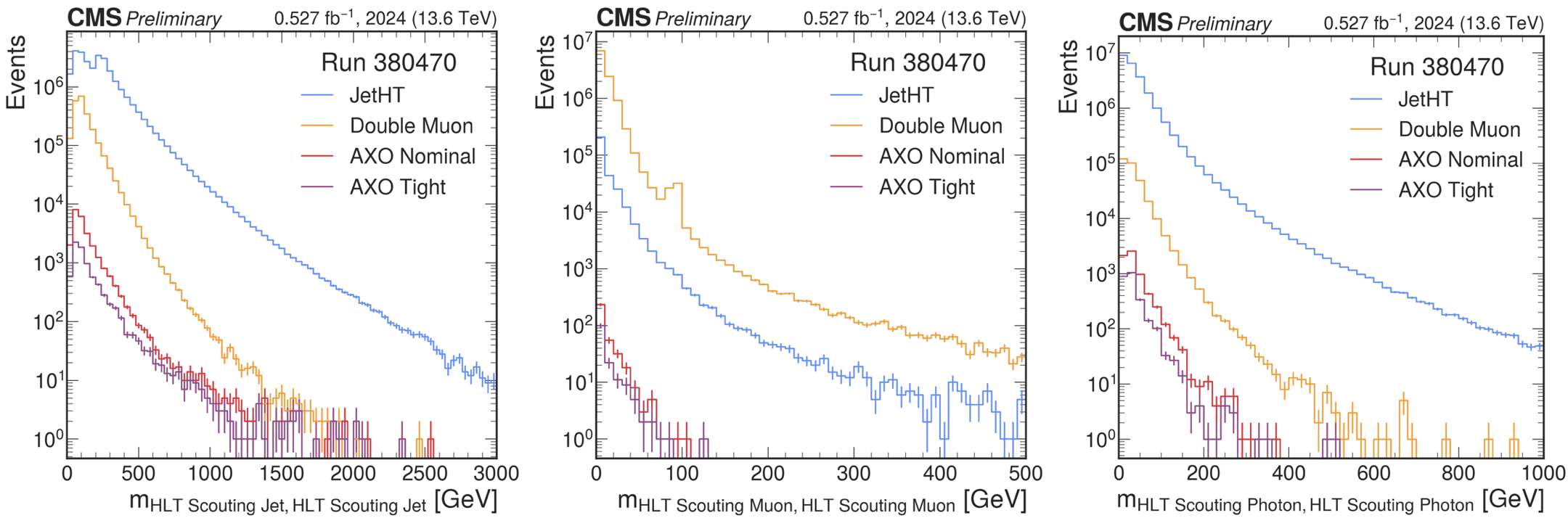} 
\caption{Invariant mass distributions of pairs of jets (left), muons (center), and photons (right) from objects reconstructed from data scouting as triggered by the AXOL1TL nominal trigger path.} 
\label{fig:mass_dist} 
\end{figure}

The events collected by AXOL1TL are further examined for the viability of BSM physics searches using this data. As a preliminary check, the invariant mass distributions of pairs of jets, electrons, and photons are studied for potential biases or sculpting in the mass distributions. As observed in Fig. \ref{fig:mass_dist}, these datasets seem to be free of any potential sculpting from trigger selections.

\section{Summary}

This paper introduces two machine learning-based anomaly detection algorithms, AXOL1TL and CICADA, for real-time event selection in the CMS Level-1 (L1) Trigger at the Large Hadron Collider. These methods are designed to identify anomalous events without predefined selection criteria, improving sensitivity to potential new physics beyond the Standard Model. AXOL1TL uses a variational autoencoder to analyze L1 trigger objects, while CICADA employs a convolutional autoencoder to process calorimeter energy deposits. Tested during CMS data-taking, AXOL1TL demonstrated stable performance and orthogonal event selection to traditional triggers, making it a promising tool for exploring novel physics signatures.



\bibliography{skeleton}

\begin{thebibliography}{14}
\providecommand{\natexlab}[1]{#1}
\providecommand{\url}[1]{\texttt{#1}}
\expandafter\ifx\csname urlstyle\endcsname\relax
  \providecommand{\doi}[1]{doi: #1}\else
  \providecommand{\doi}{doi: \begingroup \urlstyle{rm}\Url}\fi

\bibitem[Aarrestad et~al.(2021)]{Aarrestad:2021zos}
T.~Aarrestad et~al.
\newblock {Fast convolutional neural networks on FPGAs with hls4ml}.
\newblock \emph{Mach. Learn. Sci. Tech.}, 2\penalty0 (4):\penalty0 045015, 2021.
\newblock \doi{10.1088/2632-2153/ac0ea1}.

\bibitem[Chatrchyan et~al.(2008)]{cms}
S.~Chatrchyan et~al.
\newblock {The CMS Experiment at the CERN LHC}.
\newblock \emph{JINST}, 3:\penalty0 S08004, 2008.
\newblock \doi{10.1088/1748-0221/3/08/S08004}.

\bibitem[{CMS collaboration}(2024)]{dpn2024}
{CMS collaboration}.
\newblock {2024 Data Collected with AXOL1TL Anomaly Detection at the CMS Level-1 Trigger}.
\newblock 2024.
\newblock URL \url{https://cds.cern.ch/record/2904695}.

\bibitem[Coelho et~al.(2021)]{Coelho_2021}
C.~N. Coelho et~al.
\newblock Automatic heterogeneous quantization of deep neural networks for low-latency inference on the edge for particle detectors.
\newblock \emph{Nature Machine Intelligence}, 3\penalty0 (8):\penalty0 675–686, June 2021.
\newblock ISSN 2522-5839.
\newblock \doi{10.1038/s42256-021-00356-5}.
\newblock URL \url{http://dx.doi.org/10.1038/s42256-021-00356-5}.

\bibitem[Duarte et~al.(2018)]{Duarte:2018ite}
J.~Duarte et~al.
\newblock {Fast inference of deep neural networks in FPGAs for particle physics}.
\newblock \emph{JINST}, 13\penalty0 (07):\penalty0 P07027, 2018.
\newblock \doi{10.1088/1748-0221/13/07/P07027}.

\bibitem[{FastML Team}(2023)]{fastml_hls4ml}
{FastML Team}.
\newblock fastmachinelearning/hls4ml, 2023.
\newblock URL \url{https://github.com/fastmachinelearning/hls4ml}.

\bibitem[Govorkova et~al.(2022)]{govorkova2022autoencoders}
E.~Govorkova et~al.
\newblock Autoencoders on field-programmable gate arrays for real-time, unsupervised new physics detection at 40 mhz at the large hadron collider.
\newblock \emph{Nature Machine Intelligence}, 4\penalty0 (2):\penalty0 154--161, 2022.

\bibitem[Hayrapetyan et~al.(2024)]{scouting}
A.~Hayrapetyan et~al.
\newblock {Enriching the Physics Program of the CMS Experiment via Data Scouting and Data Parking}.
\newblock 3 2024.

\bibitem[Hinton et~al.(2015)Hinton, Vinyals, and Dean]{hinton}
G.~Hinton, O.~Vinyals, and J.~Dean.
\newblock {Distilling the Knowledge in a Neural Network}.
\newblock 3 2015.

\bibitem[Khachatryan et~al.(2017)]{CMS:2016ngn}
V.~Khachatryan et~al.
\newblock {The CMS trigger system}.
\newblock \emph{JINST}, 12:\penalty0 P01020, 2017.
\newblock \doi{10.1088/1748-0221/12/01/P01020}.

\bibitem[Kingma and Welling(2022)]{vae}
D.~P. Kingma and M.~Welling.
\newblock Auto-encoding variational bayes, 2022.
\newblock URL \url{https://arxiv.org/abs/1312.6114}.

\bibitem[LeCun et~al.(1989)]{cnn}
Y.~LeCun et~al.
\newblock Backpropagation applied to handwritten zip code recognition.
\newblock \emph{Neural Computation}, 1\penalty0 (4):\penalty0 541--551, 1989.
\newblock \doi{10.1162/neco.1989.1.4.541}.

\bibitem[Liu et~al.(2023)Liu, Gandrakota, Ngadiuba, Spiropulu, and Vlimant]{ryan}
R.~Liu, A.~Gandrakota, J.~Ngadiuba, M.~Spiropulu, and J.-R. Vlimant.
\newblock {Efficient and Robust Jet Tagging at the LHC with Knowledge Distillation}.
\newblock In \emph{{37th Conference on Neural Information Processing Systems}}, 11 2023.

\bibitem[Sirunyan et~al.(2020)]{CMS:2020cmk}
A.~M. Sirunyan et~al.
\newblock {Performance of the CMS Level-1 trigger in proton-proton collisions at $\sqrt{s} = 13$\,TeV}.
\newblock \emph{JINST}, 15:\penalty0 P10017, 2020.
\newblock \doi{10.1088/1748-0221/15/10/P10017}.

\end{thebibliography}

\end{document}